\documentstyle[psfig]{lamuphys}
\makeatletter
\let\chapter\hid@chapter
\makeatother

\begin{document}
\pagenumbering{arabic}
\title{High-Redshift Galaxies: The HDF and More}

\author{Alberto\,Fern\'andez-Soto\inst{1}, Kenneth M.\,Lanzetta\inst{2}, 
and Amos\,Yahil\inst{2}}

\institute{Dept. of Astrophysics and Optics, UNSW, Sydney, NSW2052, Australia 
\and Dept. of Physics and Astronomy, SUNY, Stony Brook, NY11794-3800, USA}

\maketitle

\begin{abstract}
We review our present knowledge of high-redshift galaxies,
emphasizing particularly their physical properties and the ways in
which they relate to present-day galaxies. We also present a 
catalogue of photometric redshifts of galaxies in the Hubble Deep
Field and discuss the possibilities that this kind of study
offers to complete the standard spectroscopically based surveys.
\end{abstract}

\section{Introduction}

For a long time models for galaxy formation and evolution advanced
unhampered by observations. Nowadays, however, the rapid increase in
both observational capabilities and efficiency of the selection
methods (see Steidel {\em et al.} 1995 [S95]) has converted the task
of looking for distant galaxies from one of the most difficult
challenges to an almost routine job, and large databases of high-$z$
galaxies are already being compiled (Dickinson 1998, this Volume).
Observations can now constrain the models, and this obliges us to
understand the properties of these objects in order to get a complete
image of the processes involved in the formation and evolution of
galaxies.

This study of the properties of high-$z$ galaxies is twofold.  We need
to understand the information provided by the confirmed high-$z$
galaxies. In this way we will learn about the spectral and
morphological properties of the bright end of the galaxy population,
i.e., the putative progenitors of present-day large ($L>L_*$)
galaxies. Second, the use of photometric redshift techniques applied
to deep multi-colour images (like the HDF, Williams {\em et al.}
1996) opens a wealth of statistical methods to study those faint
objects for which we cannot obtain spectroscopic information in the
near future. These studies will yield further results on the general
distribution and evolution of galaxies. The main problem for both
methods resides in the $z \approx 1-2$ range, where spectroscopic
identification of galaxies at optical wavelengths is made difficult by
the lack of spectral features.

\section{Physical Properties of the High-Redshift Galaxies}

We start with a brief review of the physical properties of high-$z$
galaxies, most of which have been selected applying colour techniques
(S95). The HDF triggered a wave of intense spectroscopic follow-up
observation (Steidel {\em et al.} 1996 [S96], Lowenthal {\em et al. }
1997 [L97], Zepf {\em et al.} 1996) that added a large number of
galaxies to the sample. Nature also provides us with a telescope
capable of amplifying the light from distant galaxies, although
plagued by geometric aberrations: {\em gravitational lensing} has been
used by several groups to discover some of the most distant known
galaxies (Trager {\em et al.} 1997 [T97], Franx {\em et al.}  1997).

\subsection{Spectral Features: Dust, Metal and Gas Content}

\begin{figure}[bt]
\centerline{\psfig{figure=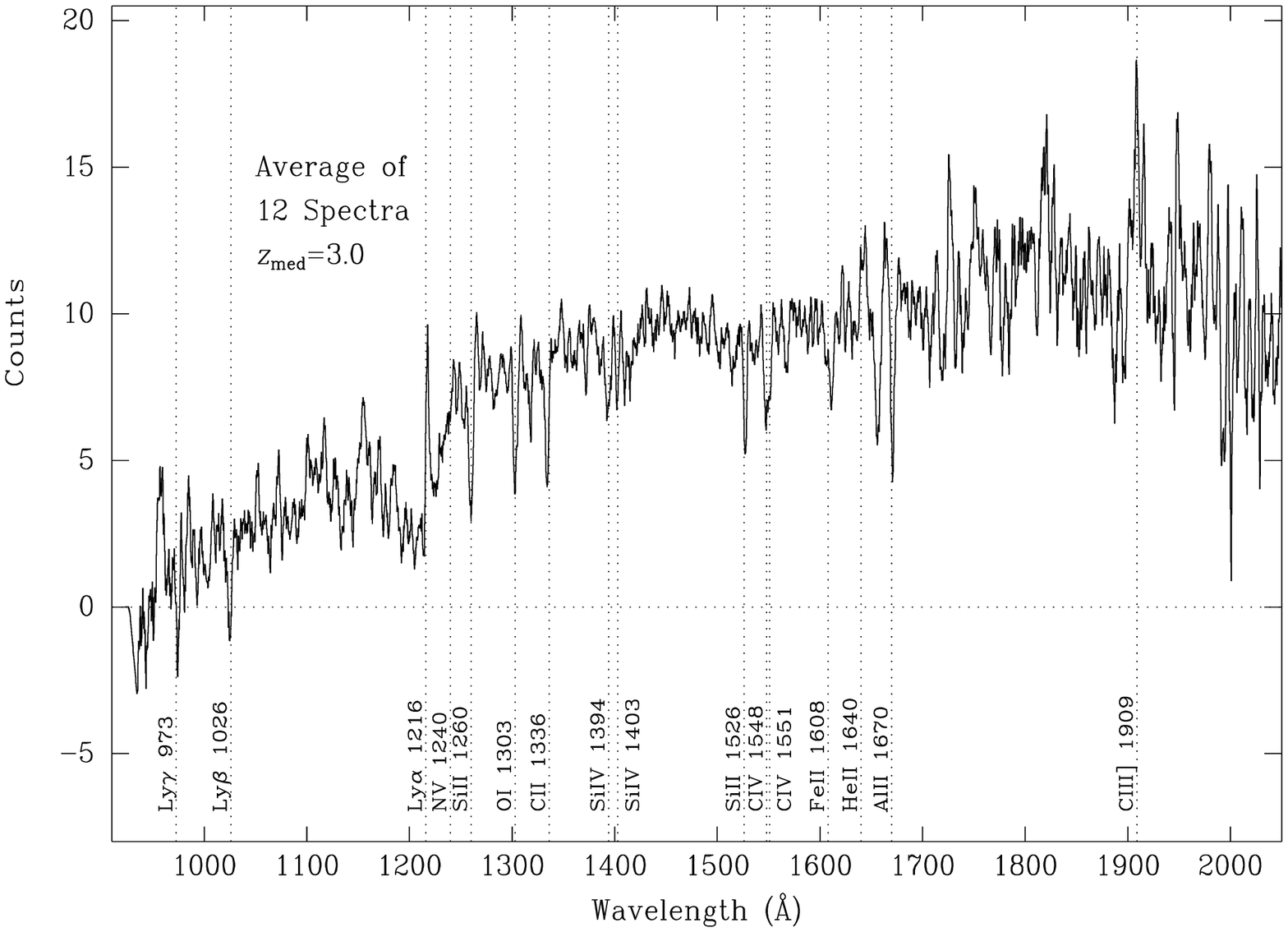,angle=-90,height=4.2cm,width=8cm}}
\caption{Average spectrum of 12 high-redshift galaxies with $z_{med}= 3.0$ 
(from L97).}
\end{figure}

High-$z$ galaxies (see Fig. 1) are characterized by a flat
continuum. Their Ly$\alpha$ emission lines vary considerably, from
weak or even absent --with superposed damped absorption profiles-- to
rest $EW$ of up to 60 \AA. All galaxies show optically thick Lyman
limits and a strong discontinuity in the continuum bluewards of
Ly$\alpha$, due to the onset of the Ly$\alpha$ forest. Stellar and
interstellar absorption lines are present, showing narrow profiles
that are weaker for high-ionization species. A detailed study of the
different emission and absorption lines and the slopes of the observed
spectra suggests that high-$z$ galaxies are low metallicity systems
($Z \approx 0.1 \ Z_{\sun}$), with high neutral gas content that
allows them to imprint damped HI absorption profiles on background
objects. The amount of dust seems to be moderate ($E(B-V) \approx
0.10$), but a better measurement is necessary in order to establish
firmer values for the extinctions, luminosities and star formation
rates.

\subsection{Morphology and Luminosity}

Most of the observed galaxies show compact cores (with half-light
radii on the order of a few kpc) surrounded by irregular asymmetric
halos (Giavalisco {\em et al.} 1996 [G96]). Although far from
homogeneous, they look more regular than the galaxies observed at $z
\approx 1$ --only one of the galaxies at $z>2$ shows the
``chain'' morphology reported by Cowie {\em et al.} (1995) to be
usual in moderate-$z$ galaxies. A joint analysis of our photometric
redshift catalogue and a morphological catalogue of galaxies in the
HDF is presented by Simon Driver in this same Volume (see also Driver
{\em et al.} 1998 [D98]). It must be remarked that we are observing
these objects in the rest-frame UV range, so passband effects are
indeed important. Direct comparison of their morphologies with those
of their low-$z$ counterparts will have to wait until high-resolution
IR imaging is available.

The total $B$-band luminosities lie in the range $1-10 L_*$, with a
strong concentration in the compact, high surface brightness
cores. The SFRs range from 1 to 50 $M_{\sun} {\rm yr}^{-1}$, although
dust extinction might increase this by a factor of perhaps 3 or even
more (Pettini {\em et al.} 1997).

\subsection{Number Densities and Clustering}

Some measured number densities are: 0.6 $\pm$ 0.2 ($R < 25.0, 3.0 < z
< 3.5$, S95), 3.2 $\pm$ 1.9 ($R < 25.3, 2.4 < z < 3.4$, S96), 6.5
$\pm$ 2.0 ($R < 25.5, 2.0 < z < 3.5$, L97). For the same ranges our
catalogue gives 0.6 $\pm$ 0.3, 3.2 $\pm$ 0.8 and 7.7 $\pm$ 1.2
galaxies per square arc minute, respectively. We also estimate that
approximately 5, 15 and 25 \% of all galaxies brighter than $AB(8140)
=$ 24, 26 and 28 respectively are at $z>2$. Evidence of large scale
structure in the distribution of high-$z$ galaxies is presented by
Mark Dickinson in this Volume.

\section{Photometric Redshifts in the HDF}

The determination of redshift via photometric methods (the ``{\em Poor
person's $z$ machine}'', as stated in Koo 1985) is a long-known
technique.  We present here some results from our catalogue, based on
$UBVIJHK$ photometry of the HDF --IR images provided by Mark Dickinson
(Dickinson {\em et al.} 1998 {\em in prep}). Full details are given
in Fern\'andez-Soto {\em et al.} 1998, ({\em in prep}). The catalogue
is essentially complete down to $AB(8140)=28$ and contains 1067
objects. 

Comparison with a sample of 106 spectroscopically determined redshifts
shows that the results are very good up to $z=1.4$ ($\Delta z_{rms} =
0.13$). At $z>2$ there is a 7\% rate of error (high-$z$ galaxies that
are assigned low redshift in our analysis), while for the rest we
obtain $\Delta z_{rms} = 0.45$. Lanzetta {\em et al.} (1997) have
shown that the number of wrong redshifts in the spectroscopic
measurements (due to misidentification of lines or operator error) are
comparable to this rate.

The advantage of this technique is the ability to estimate redshifts
for large samples of objects that are too faint to have their
redshifts spectroscopically measured ( $AB(8140) \approx 28$ vs.
$AB(8140) \approx 24$). With our sample we can estimate the $N-m-z$
distribution, the Hubble Diagram for different spectral types (see
Fig. 2), the morphological evolution of galaxies (see D98), SFR
densities (Lanzetta {\em et al.} 1998, {\em in prep}), and other
characteristics.

\begin{figure}[bt]
\centerline{\hbox{\psfig{figure=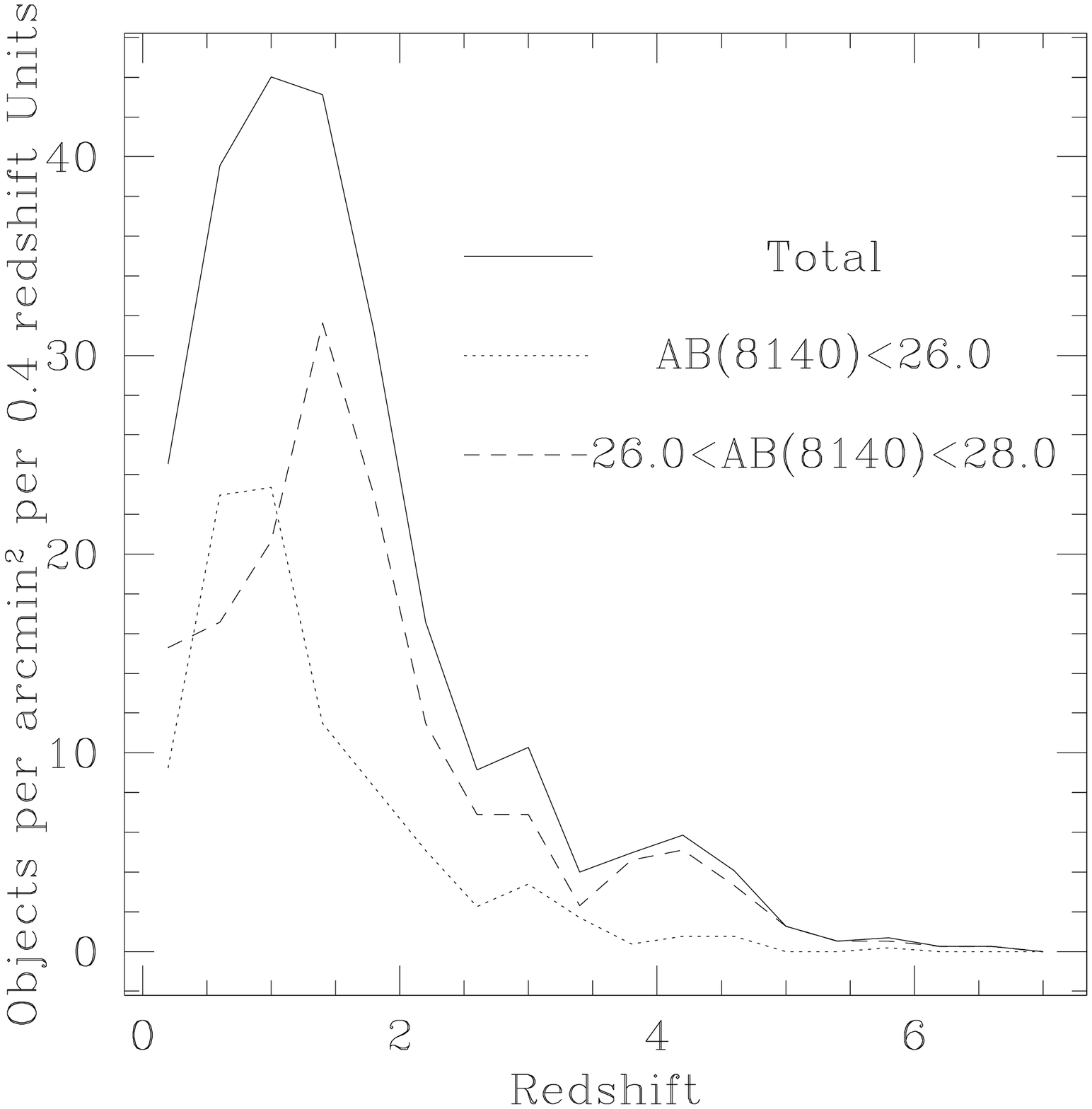,height=4.5cm,width=6cm}
                  \psfig{figure=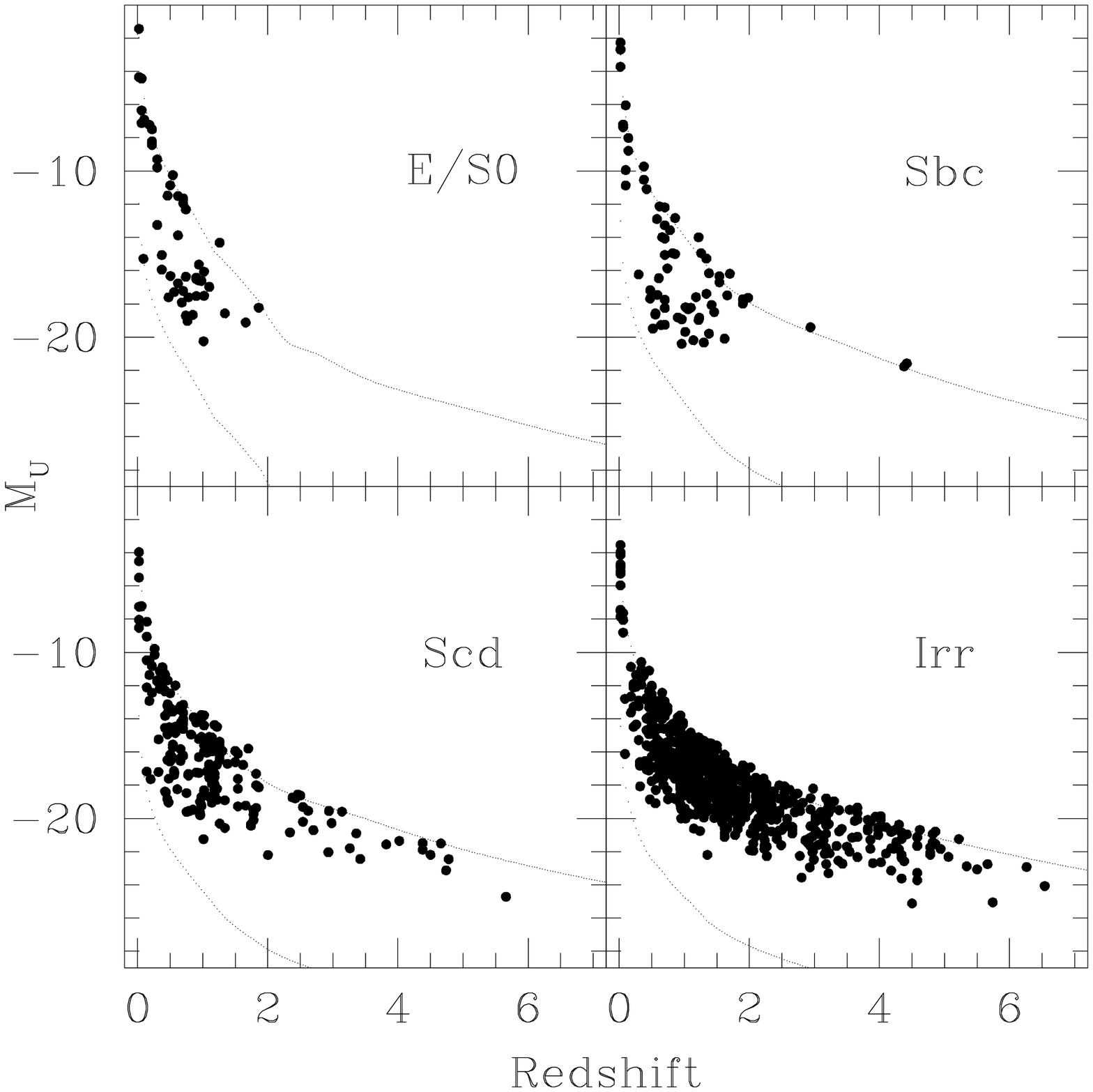,height=4.5cm,width=6cm}}}
\caption{Redshift distribution and Hubble Diagram of galaxies in the HDF.}
\end{figure}

\section{Interpretation and Conclusions}

The available data allow for different interpretations. While S95, S96
and G96 support the hypothesis that the observed high-$z$ galaxies are
the progenitors of present-day luminous galaxies at the epoch of
formation of the first stars in their spheroidal components, T97
suggests that these objects will evolve to form the Population II
components of early-type spirals.  Another interpretation (L97)
maintains that these objects represent a range of physical processes
and stages of galaxy formation and evolution rather than any
particular class of object.

While this third interpretation might be closer to reality, we are
still missing an important piece of the puzzle. Detailed IR imaging
and spectroscopy is needed in order to: a) shed light on the $z=1-2$
galaxies allowing us to constrain evolutionary models; b) obtain
images of the $z>2$ galaxies at optical rest-frame wavelengths to be
compared with their low-$z$ counterparts and; c) perform moderate
resolution spectroscopy of the $z>2$ galaxies to accurately measure
their metallicities and the importance of dust corrections.

We expect that these observations, with the support of techniques like
cosmological simulations and stellar population evolutionary models,
will lead us closer to the long-searched-for understanding of the
process by which the Universe came to be as we see it. Perhaps it is
not the moment for us to ``look deeper in the Southern Sky'', but to
look at it with different eyes.

%
%

\end{document}